\documentclass[reprint, amsmath, amssymb, aps, prl, superscriptaddress, floatfix, nofootinbib, citeautoscript, shortbibliography]{revtex4-1}
\usepackage[utf8]{inputenc}
\usepackage[T1]{fontenc}
\usepackage[parfill]{parskip}
\usepackage{siunitx}
\usepackage{blindtext}
\usepackage[colorlinks, linkcolor=blue, citecolor=blue, urlcolor=blue, breaklinks=true]{hyperref}
\usepackage{graphicx}
\usepackage{dcolumn}
\usepackage{bm}
\usepackage{color}
\usepackage{units}
\usepackage{mathtools}
\usepackage{dsfont}
\usepackage{xcolor}
\usepackage{float}

\newcommand{\vek}[1]{\boldsymbol{#1}}
\newcommand{\expec}[1]{\left\langle #1 \right\rangle}

\newcommand{\ket}[1]{\left| #1 \right\rangle}

\begin{document}

\title{Observation of the quantum Gouy phase}

\author{Markus Hiekkam\"aki}
\email{markus.hiekkamaki@tuni.fi}
\affiliation{Tampere University, Photonics Laboratory, Physics Unit, Tampere, FI-33720, Finland}

\author{Rafael F. Barros}
\affiliation{Tampere University, Photonics Laboratory, Physics Unit, Tampere, FI-33720, Finland}

\author{Marco Ornigotti}
\affiliation{Tampere University, Photonics Laboratory, Physics Unit, Tampere, FI-33720, Finland}

\author{Robert Fickler}
\email{robert.fickler@tuni.fi}
\affiliation{Tampere University, Photonics Laboratory, Physics Unit, Tampere, FI-33720, Finland}

\begin{abstract}
\noindent Controlling the evolution of a photonic quantum states is crucial for most quantum information processing and metrology tasks.
Because of its importance, many mechanisms of quantum state evolution have been tested in detail and are well understood.
However, the fundamental phase anomaly of evolving waves called the Gouy phase has not been studied in the context of elementary quantum states of light such as photon number states.
Here we outline a simple method for calculating the quantum state evolution upon propagation and demonstrate experimentally how this quantum Gouy phase affects two-photon quantum states.
Our results show that the increased phase sensitivity of multi-photon states also extends to this fundamental phase anomaly and has to be taken into account to fully understand the state evolution.
We further demonstrate how the Gouy phase can be used as a tool for manipulating quantum states of any bosonic system in future quantum technologies, outline a possible application in quantum-enhanced sensing, and dispel a common misconception related to the nature of the increased phase sensitivity of multi-photon quantum states.
\end{abstract}

\maketitle

The wave dynamics dictating the evolution of quantum states is of utmost importance in both fundamental studies of quantum systems and quantum technological applications.
For photons, the evolution of their spatial structure has been the key in a plethora of promising techniques for quantum communication \cite{sit2017high, cozzolino2019orbital}, information processing \cite{babazadeh2017high, erhard2018twisted}, simulation \cite{cardano2017detection}, and metrology \cite{hiekkamaki2021photonic}.
One particular feature of a converging wave travelling through its focus is the acquisition of an additional phase shift when compared to a collimated beam or a plane wave traveling the same distance.
This effect, which is known as the Gouy phase, was first observed and described by Gouy more than a century ago \cite{gouy1890propriete,gouy1890propagation}.
Although the phenomenon is well established and can be described through methods in physical optics \cite{baladron2019isolating,  linfoot1956phase}, the Gouy phase continues to be the topic of studies discussing its underlying physical origin by linking it to properties such as the geometry of the focus, geometric phases, and the uncertainty principle \cite{boyd1980intuitive,feng2001physical,hariharan1996gouy,visser2010origin,simon1993bargmann,baladron2019isolating,lee2020origin,subbarao1995topological,yang2006generalized}.
In addition to the continued interest aiming at providing an intuition for the phenomenon, this phase anomaly is often harnessed to realize novel tools in optics \cite{zhou2017sorting,gu2018gouy,beijersbergen1993astigmatic,whiting2003polarization}. \\
\indent Despite the Gouy phase being a general wave phenomenon, studies investigating its role in quantum state evolution have been limited to a few matter wave studies \cite{da2010indirect,da2011experimental,petersen2013measurement,ducharme2015gouy,guzzinati2013observation} and spatially separated photon pairs \cite{kawase2008observing,de2021gouy}. 
While these demonstrations utilize (locally) single quantum systems and, thus, observe the effect known for classical light waves, more complex quantum states consisting of multiple identical quantum systems, i.e., bosonic systems with multiple excitations, have not been studied before.
We term the specific phase acquired by such quantum states the quantum Gouy phase.\\

\begin{figure*}[ht]
    \centering
    \includegraphics[width = \textwidth]{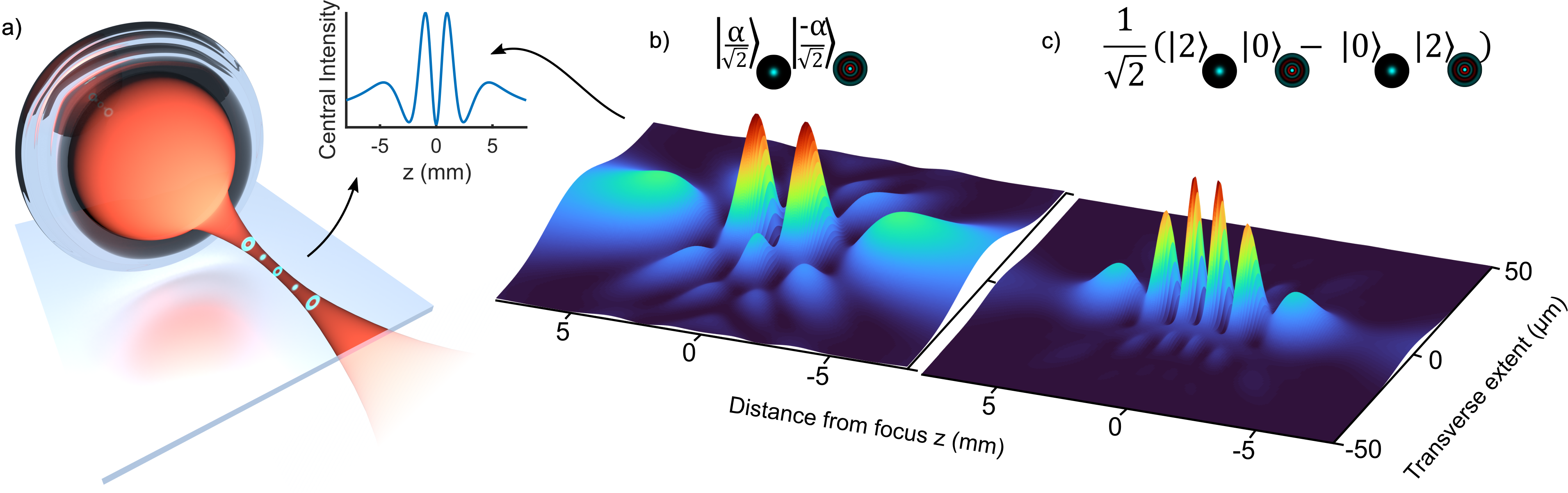}
    \caption{ 
    Observing the quantum Gouy phase through a changing on-axis interference along the propagation direction. 
    a) Conceptual image of the observation scheme. 
    The image displays the intensity structure of a superposition of a radial mode with $p' = 4$ and a Gaussian reference ($p = 0$) at different distances from the focus.
    The inset shows the intensity of the field on the optical axis.
    b) Intensity of a classical light beam prepared in the same superposition as in a), with a Gaussian waist $w_0 = 25~\mu m$ and $z_0=0$.
    c) Spatially varying two-photon probability for a two-photon N00N state prepared in the same radial modes as in a) and b).
    To make this structure visible, we post-select for cases where the two photons exist in the same position using the projection $P(x,y,z)=\vert\langle\Psi(z) \vert 2\rangle_{x,y} \vert ^2$. 
    For b) and c) the intensities/probabilities are calculated on a plane cutting through the optical axis (see a) for reference).}
    \label{fig:concept}
\end{figure*}

In general, any phase accrued by a mode of a photonic quantum system leads to a photon-number dependent phase for the quantum state.
This means that whereas a single photon or a classical field would acquire a phase $\phi$ upon propagation, when $N$-photons occupy the same mode $(\ket{N})$ the quantum state is left with $N$ times the same phase, i.e. $\exp(iN\phi)\ket{N}$ \cite{dowling2008quantum}.
This increased phase sensitivity of photon number states is utilized in so-called N00N states that have garnered popularity due to their potential in pushing the sensitivity of measurements to what is considered the absolute physical limit \cite{giovannetti2011advances}.
N00N states can be compactly expressed for two orthogonal modes $p$ and $p^\prime$ as

\begin{equation}
\label{eq:N00N}
    \ket{\Psi} = \frac{1}{\sqrt{2}}\left(\ket{N}_{p}\ket{0}_{p^\prime} - \ket{0}_{p}\ket{N}_{p^\prime}\right).
\end{equation}
Hence, the enhancement in measurement sensitivity is enabled by the phase difference between the two components being $N$ times the phase difference between the underlying modes.
More importantly, using such a N00N state configuration allows the study of the speed-up of the quantum Gouy phase compared to the classical case.\\
\indent In the present work, we describe theoretically how an $N$-photon number state evolves upon propagation and verify experimentally the speed-up of the quantum Gouy phase with two-photon N00N states through interference in the transverse structure of a bi-photon.
We further show that the quantum Gouy phase speed-up can be applied to super-resolving longitudinal displacement measurements using the quantum Fisher information (QFI) formalism and solidifying its link to the uncertainty interpretation of the Gouy phase \cite{feng2001physical}.
Finally we show that our results for N-photon states cannot be simulated by classical light with a $\lambda/N$ wavelength, demonstrating that the often-used effective de Broglie wavelength approach for multi-photon states, although useful in specific cases \cite{jacobson1995photonic,edamatsu2002measurement,walther2004broglie},  is not always accurate. 
As such, our work brings the fundamental wave feature of the Gouy phase to the quantum domain, thereby opening the path to its utilization in quantum technological applications through its unique quantum state manipulation properties.

\section{Probing the quantum Gouy phase}\label{sec2}

To observe the quantum Gouy phase of $N$-photons, an interferometric measurement scheme can be used.
We chose to use the transverse-spatial modes of paraxial light beams as the different "arms" of the interferometric scheme, where one mode acts as the required reference "arm".
More specifically, we used Laguerre-Gaussian (LG) modes, which are a family of orthogonal solutions to the paraxial wave equation in cylindrical coordinates \cite{andrews2012angular}.
In the case of a classical monochromatic field, the Gouy phase of these modes evolves as \cite{andrews2012angular}

\begin{equation}
\label{eq:Gouy}
    \Phi_G(z) = -(2p+\vert\ell\vert+1)\arctan\left(\frac{2(z-z_0)}{kw_0^2}\right),
\end{equation}
where $z$ is the propagation distance, $k$ is the wavenumber, $\ell$ is an integer giving the number of orbital angular momentum quanta per photon, $p$ is a positive integer defining the radial transverse structure of the field, $w_0$ is the beam waist defining the transverse extent of the beam at its focus, and $z_0$ gives the position of the beam focus along the optical axis.
Since the Gouy phase depends on the mode order $S = 2p+\vert\ell\vert+1$, its anomalous phase behaviour can be observed through the change of the transverse structure during propagation when the light is in a superposition of spatial modes of different mode orders \cite{da2020pattern}.
For radial modes, which are LG modes with $\ell = 0$, this change results in a varying intensity along the optical axis, as can be seen in Fig.~\ref{fig:concept}a).
Thus, to probe the quantum Gouy phase and contrast it to its classical counterpart, we study the superposition of a Gaussian reference mode ($p = 0$) and different higher order radial modes in both the classical domain and the aforementioned quantum setting, i.e., a N00N state superposition.
By measuring the change in intensity and two-photon detection rate, respectively, observed in a single mode fiber (SMF) scanned through the focus, we are able to directly observe the speed-up of the quantum Gouy phase.

\section{Theoretical evolution upon propagation}\label{sec3}

In our measurement scheme, we expect the propagation to result in a photon number dependent Gouy phase when the state $\ket{N}_p$ is translated through a focus.
To verify these expectations theoretically, we start with $N$ photons occupying a monochromatic paraxial mode at a position $z=0$, with a complex field structure $u_{\ell p}(\vek{\rho}, 0)$.
To translate the mode along the optical axis, we apply the translation operator $e^{\mathrm{i}\hat{P}_z z/\hslash}$ to the mode in the angular spectrum representation, in which the quantized mode of light can be expressed as 

\begin{equation}
    \hat{a}^\dagger_{\ell p}(0) =  \int \int F_{\ell p}(\vek{\kappa},0) \hat{a}^\dagger(\vek{\kappa}) d^2\kappa\,,
\end{equation}
where $F_{\ell p}(\vek{\kappa},0)$ represents the normalized complex amplitude of the plane wave mode with transverse wave vector $\vek{\kappa}$, and $\hat{a}^\dagger(\vek{\kappa})$ is the corresponding operator density \cite{torres2003preparation,wunsche2004quantization}.
After applying the translation operator, the mode takes the form

\begin{equation}
    \hat{a}^\dagger_{\ell p}(z) 
    = \int \int F_{\ell p}(\vek{\kappa},0) e^{-\mathrm{i}k_z(\vek{\kappa})z}\hat{a}^\dagger(\vek{\kappa}) d^2\kappa\,,
\end{equation}
which is identical to the initial mode being propagated by $z$ using the angular spectrum method (ASM) \cite{baladron2019isolating,saleh2019fundamentals}.
We thus see that the quantized mode evolves identically to a classical light field, i.e., the propagated LG mode has an identical spatial structure $u_{\ell p}(\vek{\rho}, z)$ only differing by the propagation-related change in wavefront curvature and beam radius.
Due to the beam evolving according to the ASM, we can extract the Gouy phase evolution by defining a new mode $\hat{b}^\dagger_{\ell p}(z)$ which has the structure of the field after translation, without the accumulated Gouy phase, i.e., $u_{\ell p}(\vek{\rho},z)e^{-\mathrm{i}kz}e^{\mathrm{i}\Phi_G(z)}$.
Using this new mode, we can express the mode after propagation as a single mode with a phase

\begin{equation}
    \hat{a}^\dagger_{\ell p}(z) = \hat{b}^\dagger_{\ell p}(z)e^{-\mathrm{i}\Phi_G(z)}\,.
\end{equation}
We can then simply state the Gouy phase evolution of an $N$-photon Fock state as

\begin{equation}
\label{eq:Gouy_speedup}
    \ket{N}_{\ell p;0} \rightarrow 
    e^{-\mathrm{i}Nkz-\mathrm{i}N\Phi_G(z)}\ket{N}_{\ell p;z}\,,
\end{equation}
which explicitly contains the photon number dependent Gouy phase evolution. 
For a detailed derivation, see the Supplementary.

\section{Experiment}\label{sec4}

In the experiment, we first prepared laser light in a superposition of the Gaussian reference mode and one of the higher-order radial modes.
The structuring of the laser beam was done with a single hologram on a spatial light modulator (SLM), using a holographic method commonly known as mode carving \cite{bolduc2013exact}.
After structuring, the beam was imaged one focal distance away from a 75~mm lens which performs an optical Fourier transform on the transverse structure while focusing \cite{saleh2019fundamentals}.
Since the transverse structure and its Fourier transform are identical for LG modes, the beam structure at the focus was identical to the structure carved at the SLM, up to a phase factor of $\pi$ between the superposed LG modes which needed to be accounted for with odd values of the radial index \cite{gu2018gouy,zhou2017sorting}.
To measure the Gouy phase induced change in the interference along the optical axis, we placed an SMF at the focus and moved it longitudinally using a stage with a computer controlled piezo actuator.
The laser source was a continuous-wave diode laser operating at 810~nm and the SLM used for structuring the light was wavefront corrected using the method described in Ref. \cite{jesacher2007wavefront}.
Furthermore, to get the generated modes as close as possible to the correct transverse structure at the wanted beam radius, we employed an additional Gaussian correction in the mode carving that minimized any effect of the initial Gaussian beam structure in the carved mode (see Supplementary).\\

\begin{figure}[t]
    \centering
    \includegraphics[width = \columnwidth]{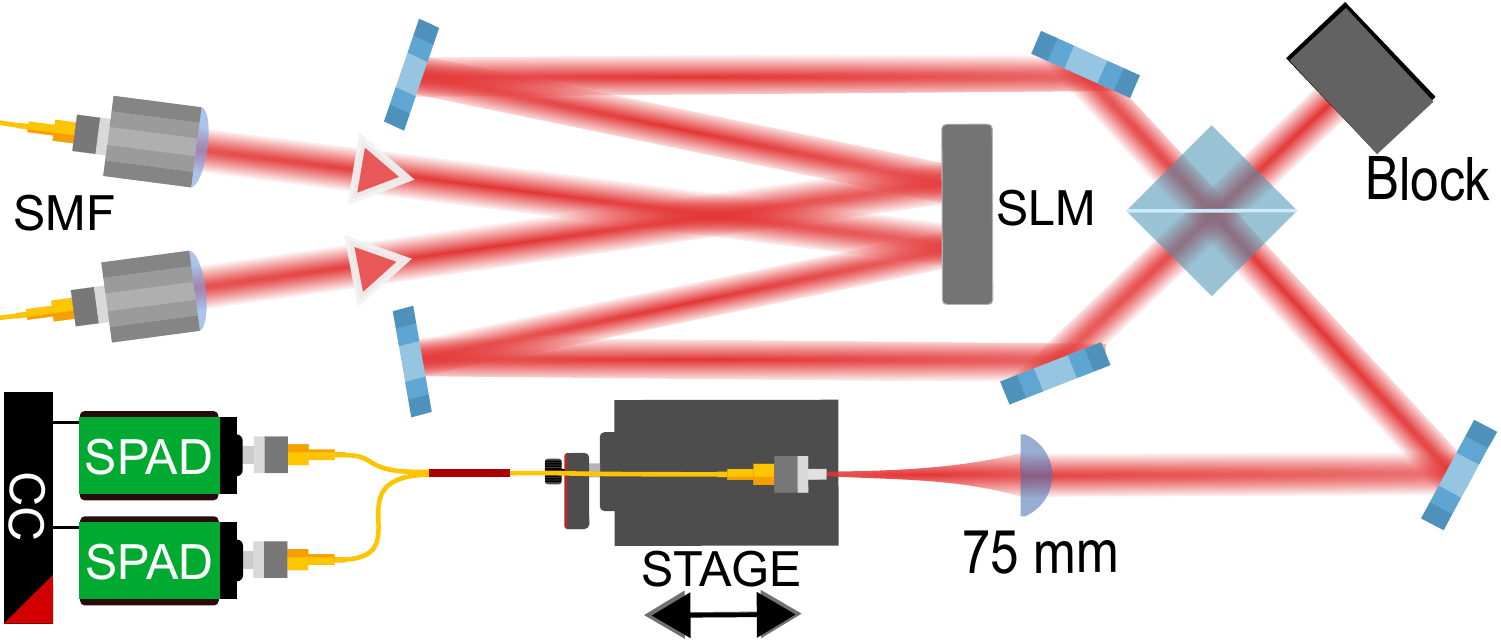}
    \caption{Simplified drawing of the experimental setup.
    Two photons with Gaussian beam profiles were sent on to separate sections of an SLM where they were independently structured into orthogonal superpositions of radial modes.
    Then these photons were probabilistically overlapped using a beamsplitter, after which they bunched into a radial mode N00N state \cite{hiekkamaki2021photonic}.
    Finally, this two-photon N00N state was focused down to a $25~\mu$m Gaussian beam waist and coupled into a SMF (with a mode field diameter of $5~\mu$m) that was scanned through the focus (from behind the focus towards the lens).
    The two-photons were then probabilistically split into two single photon avalanche diodes (SPAD) and we post-selected on both of the detectors detecting a photon at the same time using a coincidence counter (CC).
    For more details, see the main text and Supplementary.}
    \label{fig:setup}
\end{figure}

\begin{figure*}[htb]
    \centering
    \includegraphics[width = 1\textwidth]{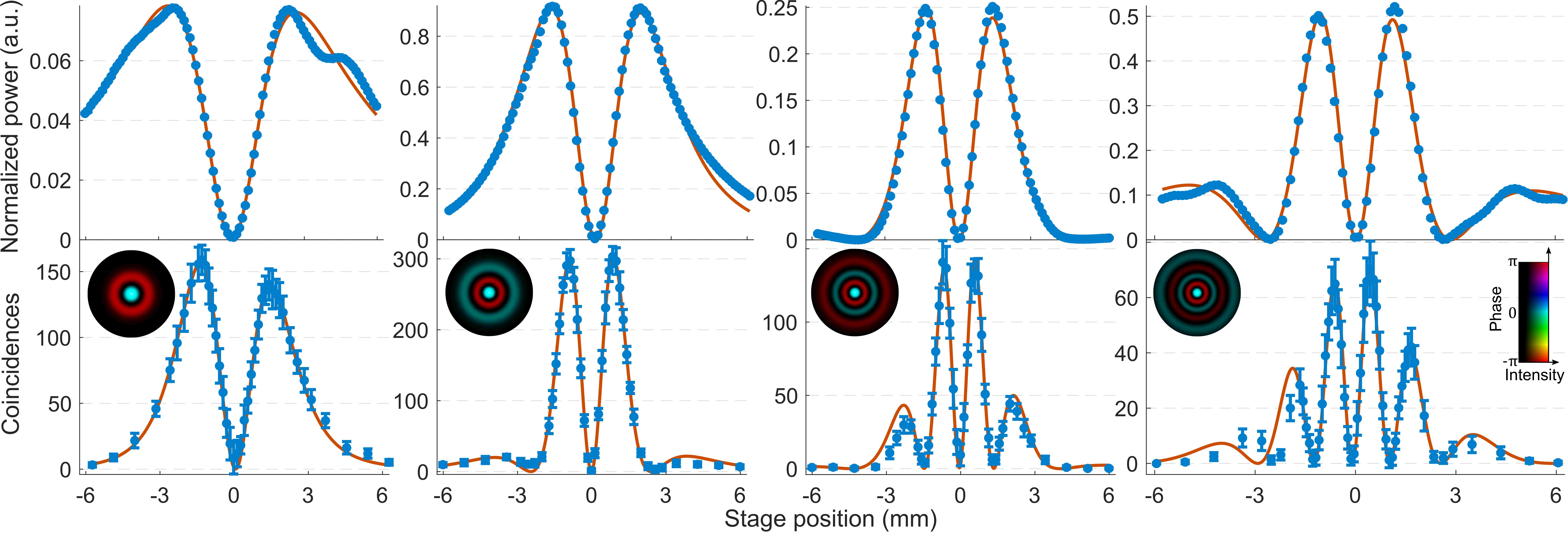}
    \caption{Comparison of on-axis interference along the propagation direction for classical light and two-photon N00N states.
    The upper row contains laser data with 100 repetitions per measurement point. 
    The lower row, shows two-photon coincidence measurements. 
    Each two-photon data point was corrected for accidental coincidence detection and measured 25 times with an integration time of 28 sec for $p^{\prime} = \{1,2,4\}$ and 24 sec for $p^{\prime} = 3$. 
    The fits are nonlinear least squares fits of the form described in the main text.
    To keep the data comparable, we aimed to keep the beam waist at $25~\mu$m in all measurements. 
    However, the fits showed that the beam waist was slightly larger for higher order modes, varying between $24.95~\mu$m and $26.81~\mu$m.
    A negative stage position labels that the SMF collecting the light was between the focus and the lens. The errorbars are $\pm1$ standard deviations and the insets show images of the corresponding radial modes with indices $p^{\prime}$.
    }
    \label{fig:Data}
\end{figure*}

\indent For a classical field, we can extract the theoretically expected measurement results simply by calculating the overlap of the Gaussian eigenmode of the SMF and the normalized transverse structure of the scalar field $u_{\mathrm{tot}}(\vek{\rho},z) = \frac{1}{\sqrt{2}}(u_{0p}(\vek{\rho},z) - e^{\mathrm{i}\theta}u_{0p^\prime}(\vek{\rho},z))$.
Thus, for laser light, the amount of laser power coupled in to the fiber is proportional to

\begin{equation}
\label{eq:laser_probab}
    P_o \propto \left\vert A_p(z) - e^{-\mathrm{i}\theta} A_{p^\prime}(z)\right\vert^2,
\end{equation}
where $A_{j}(z)$ refer to the overlap between the normalized radial mode $j$, at a distance $z$ from its focus, and the normalized Gaussian eigenmode of the fiber.
To see the Gouy phase dependence of the detection probability, the above equation can then be stated as 

\begin{equation*}
\begin{aligned}
        P_o \propto & [\vert A_p\vert^2+\vert A_{p^\prime}\vert^2\\ 
        &-2\vert A_p\vert\vert A_{p^\prime}\vert\cos(\Delta\Phi_G(z)-\theta + \phi(z))],
\end{aligned}
\end{equation*}
where the term $\phi(z)$ is an extra phase contribution from the curvature of the wavefront acquired upon propagation.
However, since the wavefront curvature is very small near the optical axis, the only significant contribution to the phase of the overlaps $A_j(z)$ comes from the Gouy phase difference $\Delta\Phi_G(z)$.
Thus, scanning the fiber through the focus results in a signal which oscillates as $\cos\left(2(p^\prime-p)\arctan\left(\frac{2(z-z_0)}{kw_0^2}\right)\right)$ underneath some envelope function caused by the $z$-dependence of the overlap functions.\\
\indent For the measurements, we kept the reference mode, i.e., a Gaussian mode with radial index $p = 0$, fixed and varied the index $p^{\prime}$ of the probe mode between 1 and 4, which lead to four different measurement scenarios with differing Gouy-phase contributions.
The measured data can be found on the top row of Fig.~\ref{fig:Data}.
The measurements follow the probability introduced above very well, which we verified by fitting curves matching Eq.~\eqref{eq:laser_probab} to the data.
In each fit, we fixed the mode field diameter of our fiber to the $5~\mu$m specified by the manufacturer and only had 4 fitting parameters: an overall scaling factor of the function, the beam waist $w_0$, focal position $z_0$, and the $z$-independent phase offset $\theta$.
The average adjusted R-squared value of the fits was $0.986$, meaning that the data corresponds well to the theoretical model. \\ 

\indent After first verifying the methods viability using a laser and showing the effect of the Gouy phase on a classical interference pattern along the optical axis, we extended the measurement scheme to observe the quantum Gouy phase.
Following the same general idea, we now generated different two-photon N00N states between a reference Gaussian mode ($p=0$) and higher order radial modes, and studied the two-photon interference pattern along the optical axis. 
To prepare such a N00N state, we first generated photon pairs through spontaneous parametric down-conversion (see Supplementary for more information) and then shaped each of the two photons individually into a well-defined superposition of the wanted radial modes using two holograms performing two different mode carvings.
Once each of the photons was structured, we directed the photons into the same beam path using a beamsplitter.
As demonstrated in \cite{hiekkamaki2021photonic}, once in the same beam path, indistinguishable photons bunch into the wanted spatial mode N00N state given in Eq.~\eqref{eq:N00N}.
A simplified sketch of the two-photon experimental setup can be seen in Fig.~\ref{fig:setup}.\\
\indent To calculate the $N$-photon coincidence probability, we project the radial mode N00N state $\ket{\Psi(z)}$ onto the state where all of the photons have been coupled successfully into the SMF $P=\vert\langle\Psi(z)\vert N\rangle_{SMF}\vert^2$.
Assuming that we produce perfectly balanced N00N states of radial modes with a phase offset $\theta$, the $N$-photon detection probability can be reduced to the form

\begin{equation}
\begin{aligned}
\label{eq:2_phot_probab}
    &P=\frac{1}{2}\left\vert A_p^{N}(z) - e^{-\mathrm{i}
    \theta} A_{p^\prime}^{N}(z)\right\vert^2.
\end{aligned}
\end{equation}
As before, we can express this coincidence probability as

\begin{equation*}
\begin{aligned}
    P=& \frac{1}{2}[\vert A_p\vert^{2N}+\vert A_{p^\prime}\vert^{2N}\\
    &-2\vert A_p\vert^{N}\vert A_{p^\prime}\vert^{N}\cos(N\Delta\Phi_G-\theta + N\phi(z))],
\end{aligned}
\end{equation*}
which is similar to the detection probability of the classical field, leading to an oscillating interference underneath some envelope function.
However, in the above equation we see the photon number dependent scaling for both the frequency of the oscillation as well as the envelope term.
Note that a probability curve with half the amplitude but same shape can also be observed for photon pairs prepared similarly without bunching.
Thus, to verify that we generate radial mode N00N states in our experiment, we prepared the two photons in the corresponding radial mode superpositions and showed that the probability of coupling both of the photons into the SMF roughly doubles when the photons are made indistinguishable in time, which is a clear signature of bunching (see Supplementary for the measured data). 
For detailed derivations of the detection probabilities, see the Supplementary. \\
\indent For the N00N state measurements, we used the same set of radial modes in superposition with the reference Gaussian mode leading to the data shown on the bottom row of Fig.~\ref{fig:Data}.
As before, the data follows very well the theoretically expected curves, verifying the above-presented equations and their described behaviours.
Fits of Eq.~\eqref{eq:2_phot_probab} to the data, with the same parameters as in the classical case, resulted in an average adjusted R-squared value of $0.951$.
The slight imperfections in the data can all be accounted for by imperfections in the alignment, imaging, the SMF eigenmode, spatial mode generation, and errors in the stage position.
Besides the errors in the stage positions, all of these can be effectively categorized as contaminations of our state space by modes not included in the theoretical analysis.
Hence, our results demonstrate that the quantum Gouy phase leads to a speed up in the accumulated phase upon propagation and also modulates the underlying envelope function.
As we will discuss next, both features shed new light on the fundamental understanding of the Gouy phase, as well as hint at quantum enhanced metrology applications.

\section{Quantum Fisher information}\label{sec5}

As the quantum Gouy phase evolves faster with a larger number of photons, one application could be super-sensitive measurements of longitudinal displacement. 
This prospect can be investigated by calculating the QFI achieved through translation, which is of the form \cite{giovannetti2011advances,demkowicz2015quantum,barbieri2022optical,polino2020photonic}

\begin{equation}
\label{eq:QFI_general}
    F_Q(\ket{\psi(z)}) = \frac{4}{\hslash^2}\Delta^2 \hat{P}_z\big\vert_\psi. 
\end{equation}
When calculating this variance for the radial mode N00N state $\ket{\Psi(z)}$ we get the QFI 

\begin{equation}
\label{eq:QFI_radial}
\begin{aligned}
    F_Q(\ket{\Psi(z)}) &= 2N\left(\Delta^2 k_z\vert_p + \Delta^2 k_z\vert_{p^\prime}\right)\\
    &+ N^2\left( \expec{k_z}_p - \expec{k_z}_{p^\prime}\right)^2,
\end{aligned}
\end{equation}
where $\Delta^2 k_z\vert_i$ and $\expec{k_z}_i$ are the variance and the average of $k_z$ for the mode $i$, respectively, calculated using the angular spectrum of the corresponding mode.
It is worth noting that the QFI does not depend on z, since the angular spectrum of a mode only acquires a phase structure upon translation.
From Eq.~\eqref{eq:QFI_radial} we can see that the second term of the QFI has Heisenberg scaling. As we show in the Supplementary, this term relates to the Gouy phase difference between modes $p$ and $p^\prime$. 
Hence, radial mode N00N states along with their quantum Gouy phase properties should be able to enhance the sensitivity of longitudinal displacement measurements.
However, although these states provide benefits such as intrinsic interferometric stability when translating the mode along $z$, the spatial extent of the modes changes, making it challenging to device a real measurement capable of saturating the QFI at any $z$.
The form of Eq.~\eqref{eq:QFI_radial} also shows that it could be possible to engineer different spatially structured quantum states to measure different physical parameters.
Due to the form of the quantum Fisher information, the key feature that needs to be optimized in such state engineering should be maximizing variance of a specific momentum of the quantum state.
For example, this would mean maximizing the variance in orbital angular momentum for rotation sensing \cite{hiekkamaki2021photonic} or linear momentum for sensing the longitudinal position \eqref{eq:QFI_general}.
See the Supplementary for the derivations of the QFI and the Fisher information calculated for the projection used in our experiment.

\section{Momentum uncertainty}\label{sec6}

In addition to showing the potential for Heisenberg scaling, there is an interesting connection between the QFI and the uncertainty interpretation of the Gouy phase which fundamentally links the potential change in the spread of the transverse momentum to the evolution of the Gouy phase \cite{feng2001physical}.
Feng and Winful also noted that a larger momentum spread of higher-order modes results in a bigger Gouy phase shift \cite{feng2001physical}.
Since the Gouy phase is increased by the photon number $N$, which is accompanied by a photon number dependent momentum spread, as can be seen in Eq.~\eqref{eq:QFI_radial}, our results make a further connection between the quantum Gouy phase and its uncertainty interpretation.
Similarly to Ref. \cite{feng2001physical}, one can further link this behaviour to a tighter spatial confinement of the photons which can be made visible, e.g. by measuring the spatial extent of the $N$-photon state as shown in Fig.~\ref{fig:concept}c).

\section{de Broglie wavelength of light}\label{sec7}

\begin{figure}[ht]
    \centering
    \includegraphics[width = \columnwidth]{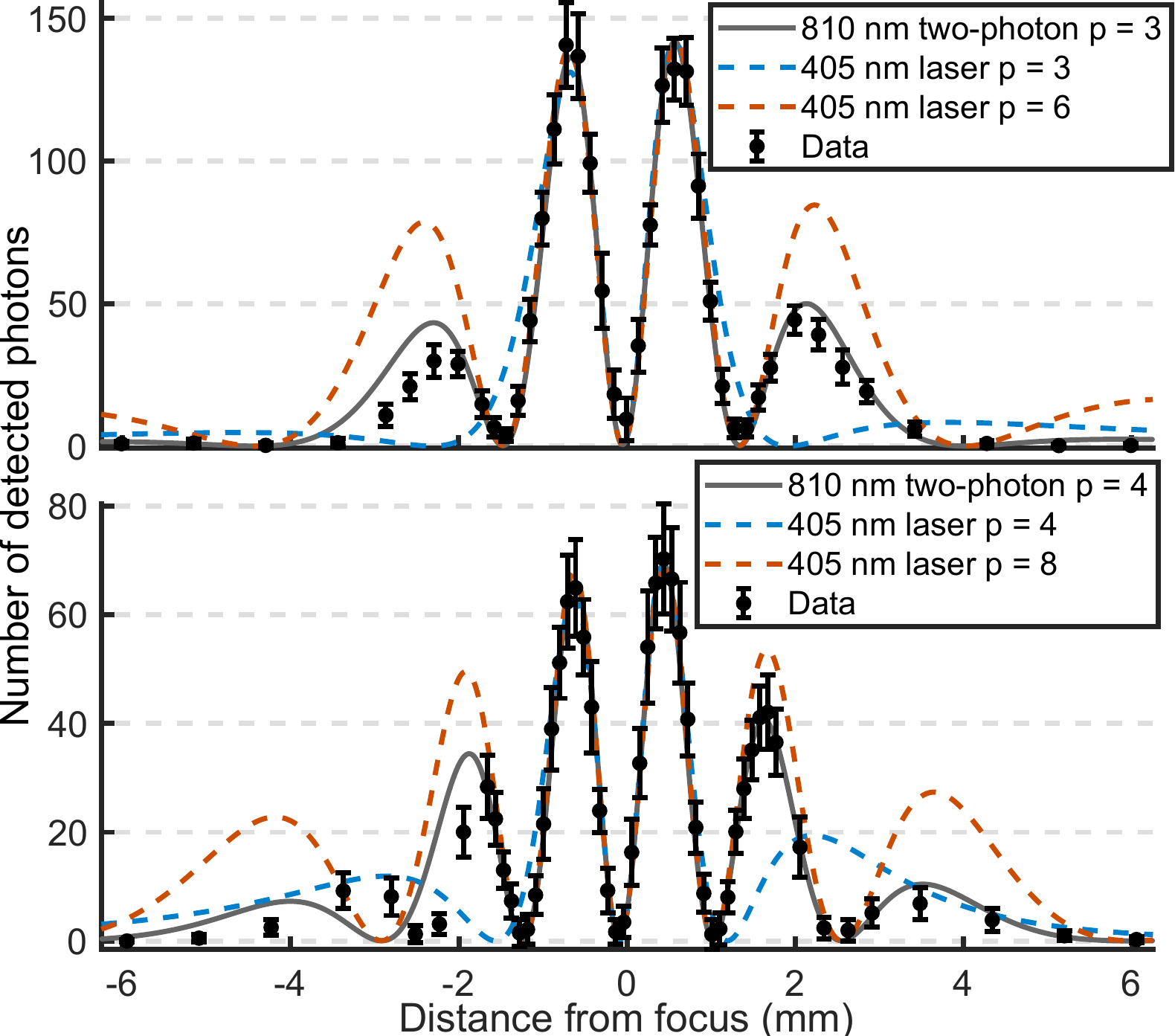}
    \caption{Comparing measured two-photon count rates with simulations of classical fields of different wavelength and mode order.
    The shown data (black dots) and its fit (grey solid line) correspond to the two-photon N00N state measurements with probe modes a) $ p^\prime=3$ and b) $p^\prime=4$.
    The dashed lines show simulated values for a 405~nm laser with two different mode orders and beam waists.
    For the blue curves, the beam radius of the 405~nm field is matched to the 810~nm mode of the photons at the focusing lens.
    For the red curves, the Rayleigh length is matched to the 810~nm mode while doubling the radial mode order $p^\prime=\{6,8\}$.
    Equation~\eqref{eq:laser_probab} with a scaling factor was used to calculate the curves for the classical 405~nm beam and the SMF mode field diameters were scaled to match the change in mode radius (i.e. $w_\mathrm{fiber}^{405} = \frac{w_0^{405}}{w_0^{810}} w_\mathrm{fiber}^{810}$).
    The curves match well near the focus.
    However, the blue curve does not exhibit the same fringe pattern and the red curve has a larger relative amplitude outside the focal region.
    Hence, the quantum Gouy phase behaviour cannot be exactly reproduced by simply changing the wavelength and mode order.}
    \label{fig:wavelength_compare}
\end{figure}

Finally, our results show that the behaviour of a two-photon N00N state cannot be replicated simply by switching to a classical field with half the wavelength.
The difference is clear if we note that the Gouy phase has a nonlinear dependence on the wavenumber which means that simply ascribing an effective de Broglie wavelength $\lambda/N$ to the $N$-photon state does not produce the correct quantum Gouy phase.
This is in contrast to the phase accrued by a non-converging field upon propagation and arguments discussed in such a context \cite{jacobson1995photonic,edamatsu2002measurement,walther2004broglie}.
In order to investigate this fundamental difference in more detail, we have plotted in Fig.~\ref{fig:wavelength_compare} the measured data for two radial mode N00N states, along with overlap curves calculated for classical 405~nm modes with two different mode orders and waists.
From these comparisons we see that the effect is not reproduced by a simple switching of the wavelength or doubling of the mode order.

Based on the comparison in Fig.~\ref{fig:wavelength_compare} and Eq.~\eqref{eq:Gouy_speedup} the only exact description of the $N$-photon Fock state evolution seems to be that it evolves as the underlying mode, taken to the power of $N$.
Although, doubling the mode order and halving the wavelength seems to replicate quite well the shown two-photon behaviour.
Since the state evolves as the mode taken to poower $N$, this evolution of the $N$-photon quantum state results in a more rapid phase change and tighter confinement of the $N$-photon.
Both of these features have been taken advantage of in different studies and experiments.
Either in the form of N00N-state super-resolution measurements \cite{slussarenko2017unconditional,dowling2008quantum} or in increasing the confinement \cite{wildfeuer2009resolution}.

\section{Conclusion}\label{sec8}

In summary, here we have verified in theory and experiment that the increased phase sensitivity of multi-photon quantum states also extends to the fundamental phase anomaly of converging waves called the Gouy phase. 
We have shown through single-path interferometric measurements along the optical axis, that two-photon N00N states experience twice the Gouy phase when traveling through a focus.
Since the Gouy phase is a fundamental feature of converging waves, our results should apply broadly to quantum states of any bosonic system.
Moreover, as the Gouy phase is an important factor in systems such as optical cavities \cite{ackemann2001gouy,wildfeuer2009resolution}, and a powerful tool in various applications such as mode sorters and mode converters \cite{gu2018gouy,zhou2017sorting,beijersbergen1993astigmatic}, our results can be widely utilized in applications in quantum optics and quantum information science.
In addition to providing a tool for quantum state manipulation, we showed that our results allow Heisenberg-limited scaling in measurements of the longitudinal displacement and, as such, might inspire new superresolution measurement schemes. \\
\indent Besides these possible technological applications, we have linked the speed-up of the Gouy phase in the quantum domain to an increased spread in the momentum of an $N$-photon state.
Hence, our results show that the uncertainty interpretation of the phase anomaly \cite{feng2001physical} holds true in the quantum domain.
Finally, due to the nonlinear relation between the Gouy phase and the wavenumber, our results unambiguously demonstrate that an $N$-photon state cannot be rigorously modelled by using a classical field with a wavelength $\lambda/N$.
However, our results suggest that an additional $N$-fold increase in the mode order can reproduce the effect of the quantum Gouy phase when the beam Rayleigh lengths are matched.
This hints at a possible link between an $N$-photon state and the $N$th harmonic of a classical field, which introduces an increase of the mode order and decrease of the beam waist, in addition to doubling the frequency.
Thus, our study not only outlines possible applications using the quantum features of spatially structured photons, it also sheds new light on the fundamental understanding of the Gouy phase, a property intrinsic to all systems described by converging or diverging waves.

\bibliography{Biblio_30032022.bib}
\begin{footnotesize}
\noindent\textbf{Supplementary material:} Accompanying Supplementary includes detailed derivations and experimental details. \\
\textbf{Acknowledgements}\\
The authors thank Fr\'ed\'eric Bouchard and Shashi Prabhakar for fruitful discussions.
\textbf{Funding:}
The authors acknowledge the support of the Academy of Finland through the Competitive Funding to Strengthen University Research Profiles (decision 301820), (Grant No. 308596), and the Photonics Research and Innovation Flagship (PREIN - decision 320165). 
MH acknowledges support from the Doctoral School of Tampere University and the Magnus Ehrnrooth foundation through its graduate student scholarship.
RF acknowledges support from the Academy of Finland through the Academy Research Fellowship (Decision 332399).
The authors declare no competing interests.
\textbf{Author Contributions:} MH and RF conceived and designed the experiment. MH constructed and performed the experiment, and processed the data. RFB and RF supervised and assisted at every stage of the study. MH, RFB, and MO derived the theoretical framework. MH, RFB, and RF wrote the manuscript. All the authors edited and proofed the manuscript.
\textbf{Competing interests:} The authors declare no competing interests.
\textbf{Data availability:} Data is available upon reasonable request.

\end{footnotesize}
\end{document}


\title{Supplementary material to: \\ Observation of the quantum Gouy phase}

\author{Markus Hiekkam\"aki}
\email{markus.hiekkamaki@tuni.fi}
\affiliation{Tampere University, Photonics Laboratory, Physics Unit, Tampere, FI-33720, Finland}

\author{Rafael F. Barros}
\affiliation{Tampere University, Photonics Laboratory, Physics Unit, Tampere, FI-33720, Finland}

\author{Marco Ornigotti}
\affiliation{Tampere University, Photonics Laboratory, Physics Unit, Tampere, FI-33720, Finland}

\author{Robert Fickler}
\email{robert.fickler@tuni.fi}
\affiliation{Tampere University, Photonics Laboratory, Physics Unit, Tampere, FI-33720, Finland}
\maketitle
\onecolumngrid

\section*{Derivation of the quantum Gouy phase}\label{sec1}

The quantum Gouy phase is the phase acquired by a Fock state of a paraxial mode upon propagation. To derive this phase, let us consider the transformation of the mode corresponding to the operator \cite{torres2003preparation, wunsche2004quantization}

\begin{equation}
    \hat{a}^\dagger_{\ell p}(0) = \int \int u_{\ell p}(\vek{\rho},0) \hat{a}^\dagger(\vek{\rho}) d^2 \rho \,,
\end{equation}
upon a translation of $z$ along the optical axis

\begin{equation}
    \hat{a}^\dagger_{\ell p}(z)= e^{-\textrm{i}\hat{P}_z z/\hslash}\hat{a}^\dagger_{\ell p}(0)e^{\textrm{i}\hat{P}_z z/\hslash}\,,
    \label{eq:translation-raw}
\end{equation}
where $\hat{a}^\dagger_{\ell p}(z)$ represents the transverse mode at any plane $z$ characterized by the integers $\ell$ and $p$, $\vek{\rho}$ contains the transverse coordinates, $\hat{a}^\dagger(\vek{\rho})$ is an operator density, and $u_{\ell p}(\vek{\rho},0)$ is the normalized structure of the mode.
The operator $\hat{P}_z$ is the longitudinal component of the linear momentum operator, defined for a monochromatic field in the plane wave basis as 

\begin{equation}
\hat{P}_z = \int \int \hslash k_z(\vek{\kappa}_j)\hat{a}^{\dagger}(\vek{\kappa}_j) \hat{a}(\vek{\kappa}_j) d^2 \kappa_j\,,
\label{eq:momentum-operator}
\end{equation}
where $k_z(\vek{\kappa}_j)$ is the longitudinal wave vector for the mode $j$, which depends on the transverse wave vector $\vek{\kappa}_j$ as $k_z(\boldsymbol{\kappa}_j)=\sqrt{k^2-\norm{\vek{\kappa}_j}^2}$, with $k$ being the wave number, common to all of the plane wave modes $\hat{a}^{\dagger}(\vek{\kappa}_j)$.
To operate on our mode with this unitary, we initially express our mode in the plane wave basis as

\begin{equation}
    \hat{a}^\dagger_{\ell p}(0) =  \int \int F_{\ell p}(\vek{\kappa},0) \hat{a}^\dagger(\vek{\kappa}) d^2\kappa\, ,
\end{equation}
where $F_{\ell p}(\vek{\kappa},0)$ is the normalized angular spectrum of the mode $\ell p$ at $z = 0$.
Now we can restate Eq.~\eqref{eq:translation-raw} 

\begin{equation}
    \hat{a}^\dagger_{\ell p}(z) =  \int \int F_{\ell p}(\vek{\kappa},0) e^{-\textrm{i}\hat{P}_z z/\hslash} \hat{a}^\dagger(\vek{\kappa})e^{\textrm{i}\hat{P}_z z/\hslash} d^2\kappa,
\end{equation}
then using the Baker-Hausdorff lemma, we can finally obtain

\begin{equation}
\label{eq:AS-propagated}
    \hat{a}^\dagger_{\ell p}(z) = \int \int F_{\ell p}(\vek{\kappa},0) e^{-\textrm{i}k_z(\vek{\kappa})z}\hat{a}^\dagger(\vek{\kappa}) d^2\kappa.
\end{equation}
The form of Eq.~\eqref{eq:AS-propagated} is now identical to a mode with an angular spectrum $F_{\ell p}(\vek{\kappa},0)$ propagated by a distance $z$ using the angular spectrum method (ASM) \cite{baladron2019isolating,saleh2019fundamentals}.
Since the ASM accounts for every aspect of paraxial beam propagation, including the Gouy phase and the plane wave phase $e^{-\textrm{i}kz}$, we can extract the Gouy phase out of the translated mode by initially expressing the translated mode in real space using the complex field distribution

\begin{equation}
\begin{aligned}
     \hat{a}^\dagger_{\ell p}(z) &= \int \int F_{\ell p}(\vek{\kappa},0) e^{-\textrm{i}k_z(\vek{\kappa})z}\hat{a}^\dagger(\vek{\kappa}) d^2\kappa
     = e^{-\textrm{i}kz}\int \int u_{\ell p}(\vek{\rho},z)\hat{a}^\dagger(\vek{\rho}) d^2 \rho \,,
\end{aligned}
\end{equation}
where $u_{\ell p}(\vek{\rho},z)$ is the normalized transverse field for the mode, calculated classically through the ASM.
We can then choose an artificial set of orthogonal spatial modes $\hat{b}^\dagger_{\ell p}(z)$ which have exactly the structure of the $\ell p$ modes at the position $z$, without the Gouy phase, i.e., $u^\prime_{\ell p}(\vek{\rho},z) = u_{\ell p}(\vek{\rho},z)e^{-\textrm{i}kz}e^{\mathrm{i}\Phi_G}$.
Expressing the position basis mode density in this new mode basis $\hat{a}^\dagger(\vek{\rho}) = \sum_{\ell^\prime p^\prime} u^{\prime *}_{\ell^\prime p^\prime}(\vek{\rho},z) \hat{b}^\dagger_{\ell^\prime p^\prime}(z) $ the translated mode takes the form

\begin{equation}
    \hat{a}^\dagger_{\ell p}(z) = \sum_{\ell^\prime p^\prime} e^{-\mathrm{i}\Phi_G} \hat{b}^\dagger_{\ell^\prime p^\prime}(z) \int \int u^*_{\ell^\prime p^\prime}(\vek{\rho},z)u_{\ell p}(\vek{\rho},z) \, d^2 \rho  \,.
\end{equation}
Due to the orthonormality of the chosen spatial mode basis \cite{andrews2012angular}, the above integral reduces to

\begin{equation}
    \hat{a}^\dagger_{\ell p}(z) = \sum_{\ell^\prime p^\prime} \delta_{\ell^\prime \ell}\delta_{p^\prime p}e^{-\mathrm{i}\Phi_G} \hat{b}^\dagger_{\ell^\prime p^\prime}(z) = e^{-\mathrm{i}\Phi_G} \hat{b}^\dagger_{\ell p}(z) \, .
\end{equation}
Hence, we can then state the evolution of an N-photon Fock state in the spatial mode $\ell p$ as

\begin{equation}
\begin{aligned}
    \ket{N}_{\ell p;0} = \frac{\left(\hat{a}^\dagger_{\ell p}(0)\right)^N}{\sqrt{N!}}\ket{0} \rightarrow & \frac{\left(\hat{b}^\dagger_{\ell p}(z)e^{-\mathrm{i}(kz+\Phi_G)}\right)^N}{\sqrt{N!}}\ket{0} 
    = e^{-\textrm{i}Nkz-\mathrm{i}N\Phi_G}\ket{N}_{\ell p;z}\,,
\end{aligned}    
\end{equation}
showing that upon translation, any phase accumulated by the mode results in N-times the same phase being accumulated by the N-photon Fock state.\\
\indent Interestingly, from the fact that any arbitrary mode $\hat{b}^\dagger_{z}$ can be chosen, one can infer that any changes in the amplitude of the field during propagation is also magnified N-times.
Similarly to the N-times increase in the phase however, in order to observe this N-fold change in the amplitude the measurement needs to be chosen in a manner that displays this change.
One example of such a measurement is the post-selection on both of the photons existing in the same spatial position which manifests as an increased confinement in Fig. 1c) of the main text.
Similarly, the same can be seen in the measured data in Fig. 3 as a narrowing of the envelope within which the fringes are observed. 
As a result, these changes could be summarized as the state experiencing any change in the mode to the power of N, which is in conflict with the effective de Broglie wavelength interpretation whenever the amplitude changes or a phase is acquired that has a nonlinear dependence on the wavenumber.

\section*{Derivation of the measurement probability}\label{sec2}

To model the expected measurement signal we calculate the probability of N photons coupling into a single mode fiber (SMF) from a radial mode N00N state.
We start off with the radial mode N00N state at position z

\begin{equation}
    \ket{\Psi(z)} = \frac{1}{\sqrt{2}}\left(\ket{N}_{0 p;z}\ket{0}_{0 p^\prime;z} - e^{\mathrm{i}\theta}\ket{0}_{0 p;z}\ket{N}_{0 p^\prime;z} \right) \, ,
    \label{eq:N00N-state}
\end{equation}
where $\theta$ is a constant phase offset between the two terms.
By decomposing the eigenmode of the SMF into a superposition of LG modes at a distance z from the beam waist

\begin{equation}
    \hat{a}^\dagger_{f} = A_{p}(z)\hat{a}^\dagger_{0p}(z)+A_{p^\prime}(z)\hat{a}^\dagger_{0p^\prime}(z)+\dots + A_{nq}(z)\hat{a}^\dagger_{nq}(z) + \dots \, ,
\end{equation}
we can calculate the probability of N photons coupling into the SMF using

\begin{equation}
\label{eq:overlap_1}
\begin{aligned}
    P =& \frac{1}{N!}\vert \langle\Psi(z)\vert \hat{a}_f^{\dagger N}\vert 0\rangle\vert ^2 \\
    =&\frac{1}{2N!} \Big\vert \left(\bra{N}_{0 p;z}\bra{0}_{0 p^\prime;z} - e^{-\mathrm{i}\theta}\bra{0}_{0 p;z}\bra{N}_{0 p^\prime;z}\right) \\
    &\left(A_{p}(z)\hat{a}^\dagger_{0p}(z)+A_{p^\prime}(z)\hat{a}^\dagger_{0p^\prime}(z)+\dots\right)^N\ket{0}\Big\vert ^2 \, .
\end{aligned}
\end{equation}
Note that in the above equations, using the normalized transverse structures, the overlaps are defined as 

\begin{equation}
\label{eq:overlap_2}
    A_{\ell p}(z) = \int \int u^*_{\ell p}(\vek{\rho},z)u_{\textrm{SMF}}(\vek{\rho}) d^2 \rho \,,
\end{equation}
which also includes the Gouy phase of each mode (note that for $A_{p}(z)$, $\ell=0$).
From the overlap calculation in Eq.~\eqref{eq:overlap_1}, only the states with N photons in either mode $0p$ or $0p^\prime$ survive and the above equation simplifies to

\begin{equation}
\begin{aligned}
    P = & \frac{1}{2}\Big\vert \left(\bra{N}_{0 p;z} - e^{-\mathrm{i}\theta}\bra{N}_{0 p^\prime;z}\right)\\ &\left(A^{N}_{p}(z)\ket{N}_{0 p;z} + A^{N}_{p^\prime}(z)\ket{N}_{0 p^\prime;z}\right)\Big\vert ^2 \\
    =& \frac{1}{2} \left\vert  A^{N}_{p}(z) - e^{-\mathrm{i}\theta}A^{N}_{p^\prime}(z) \right\vert ^2 \, .
\end{aligned}
\end{equation}
To get a more intuitive expression of this equation, we can further state $A_{i}(z) = e^{\mathrm{i}(\Phi_G(z)+\phi(z))}\vert A_{i}(z)\vert  $, where $\phi(z)$ is small since the SMF only probes the phase difference close to the optical axis where the phase of the overlap is mostly dependent on the Gouy phase and not the changing wavefront curvature.
By then defining the Gouy phase difference as $\Delta \Phi_G(z) = \Phi^\prime_G -\Phi_G$ ($\Phi_G$ and $\Phi^\prime_G$ correspond to Gouy phases acquired by LG modes of different mode order $S = 2p+1$ and $S^\prime = 2p^\prime+1$) we can write the probability as

\begin{equation}
\label{eq:quantum_Prob}
\begin{aligned}
    P =& \frac{1}{2} \left\vert  \vert A_{p}(z)\vert ^N - e^{\mathrm{i}(N\Delta \Phi_G(z)-\theta + N\phi(z))}\vert A_{p^\prime}(z)\vert ^N \right\vert ^2 \\
      =& \frac{1}{2}\Big[\vert A_p(z)\vert ^{2N}+\vert A_{p^\prime}(z)\vert ^{2N}\\
      &-2\vert A_p(z)\vert ^{N}\vert A_{p^\prime}(z)\vert ^{N}\cos(N\Delta\Phi_G(z)-\theta + N\phi(z))\Big]\, .
\end{aligned}
\end{equation}
From the form of this equation, it is clear that the probability amplitude oscillates according to $\cos(2N(p^\prime-p) \times \text{atan}(z/z_R))$, meaning that the oscillations get less frequent the further away we go from $z = 0$. 
In addition to this, the oscillation happens inside an envelope function defined by the overlaps $\vert A_p\vert $ and $\vert A_{p^\prime}\vert $.

\indent To derive a similar expected signal for the case of a classical monochromatic field with the normalized transverse structure $u_{\textrm{tot}}(\vek{\rho},z) = \frac{1}{\sqrt{2}}(u_{0p}(\vek{\rho},z) - e^{\mathrm{i}\theta}u_{0p^\prime}(\vek{\rho},z))$ being coupled into the SMF, we can calculate the power of the light in the SMF as an overlap between the structure of the incident field and the normalized eigenmode of the SMF

\begin{equation}
\label{eq:classical_Prob}
\begin{aligned}
        P_o &\propto \left\vert \int \int\left(u_{0p}(\vek{\rho},z) - e^{\mathrm{i}\theta}u_{0p^\prime}(\vek{\rho},z)\right)^*  u_{SMF}(\vek{\rho}) d^2 \rho \right\vert ^2 \\
        & =  \left\vert \int \int u^*_{0p}(\vek{\rho},z)u_{SMF}(\vek{\rho}) d^2 \rho - e^{-\mathrm{i}\theta}\int \int u^*_{0p^\prime}(\vek{\rho},z)u_{SMF}(\vek{\rho})   \right\vert ^2 \\
        &= \left\vert  A_p(z) - e^{-i\theta} A_{p^\prime}(z) \right\vert ^2 \\
        &= \left[\vert A_p\vert ^2+\vert A_{p^\prime}\vert ^2-2\vert A_p\vert \vert A_{p^\prime}\vert \cos(\Delta\Phi_G-\theta+\phi(z))\right] \, ,
\end{aligned}
\end{equation}
where the overlaps $A_{i}(z)$ correspond to the same overlaps calculated in Eq.~\eqref{eq:overlap_2} and the same assumptions can be made about the overlaps near the optical axis.

\section*{Overlap between the eigenmode of a fiber and a Laguerre-Gaussian mode} \label{sec3}

To speed up the data processing, we analytically derived the overlap of a monochromatic p-mode and a Gaussian mode corresponding to the eigenmode of the SMF

\begin{equation}
\label{eq:OV_initial}
    A_p(z) = \int \int u^*_{0p}(\vek{\rho},z)u_{\textrm{SMF}}(\vek{\rho}) d^2 \rho \,,
\end{equation}
where the normalized p-mode is defined according to \cite{kawase2008observing}

\begin{equation}
\begin{aligned}
  u_{0p}(r,\varphi,z) = & \sqrt{\frac{2}{\pi}} \frac{1}{w(z)}  \exp\left(-\frac{r^2}{w^2(z)}\right) L_p\left(\frac{2 r^2}{w^2(z)}\right)
  \exp \left(\textrm{i}\left[ \frac{r^2k}{2R} - (2p+1)\arctan\left(\frac{z}{z_R}\right) \right]\right) \, , 
\end{aligned}
\end{equation}
where $r$ is the radial coordinate, $\varphi$ is the azimuthal coordinate, $w(z) = w_0\sqrt{1+[(z-z_0)/z_R]^2}$ is the beam radius, $L_n(x) = \sum_{j=0}^n \binom{n}{j} \frac{(-1)^j}{j!}x^j$ is the Laguerre polynomial, $w_0$ is the beam waist, $R = z\left(1+[z_R/(z-z_0)]^2\right)$ is the curvature radius, $z_R = \frac{kw_0^2}{2}$ is the Rayleigh length, and $k$ is the wave number.
The normalized eigenmode of the fiber can be similarly defined as

\begin{equation}
    u_{SMF}(r,\varphi) = \sqrt{\frac{2}{\pi}} \frac{1}{w_f} \exp\left(-\frac{r^2}{w_f^2}\right) \, .
\end{equation}
For simplicity we have set $z_0 = 0$ and insert the above definitions into equation~\eqref{eq:OV_initial}

\begin{equation*}
\begin{aligned}
    A_p(z) = & \frac{2}{\pi} \frac{1}{w_f w(z)} \exp \left(\textrm{i}(2p+1)\arctan\left[\frac{z}{z_R}\right]\right) \times 
    \int_0^\infty\int_0^{2\pi} r \exp\left(-r^2\left[\frac{1}{w_f^2} + \frac{1}{w^2(z)} + \textrm{i}\frac{k}{2R}\right]\right)\times \\
    &\sum_{j=0}^p \binom{p}{j} \frac{(-1)^j}{j!}\left(\frac{2r^2}{w^2(z)}\right)^j d\varphi dr \, .
\end{aligned}
\end{equation*}
If we then define $B(z) = \frac{w(z)}{w_f} \exp \left(\textrm{i}(2p+1)\arctan\left[\frac{z}{z_R}\right]\right)$, $C(z) = \frac{w^2(z)}{2}\left(\frac{1}{w_f^2} + \frac{1}{w^2(z)} + \textrm{i}\frac{k}{2R(z)}\right)$, and $x(z) = \frac{2r^2}{w^2(z)}$, we can simplify the above equation to

\begin{equation}
    A_p(z) = B(z) \int_0^\infty \exp\left[- C(z) x\right] \sum_{j=0}^p \binom{p}{j} \frac{(-1)^j}{j!}x^j dx \, .
\end{equation}
We can then use the identity $\int_0^\infty y^n \exp\left[-a y\right] dy = \frac{n!}{a^{n+1}}$, which holds when $Re(a) > 0$ and $n \in \mathbb{N}$, to arrive at the expression

\begin{equation}
    A_p(z) = B(z) \left[\sum_{j=0}^p \binom{p}{j} \frac{(-1)^j}{C^{j+1}(z)]}\right] \, ,
\end{equation}
which we then used for calculating the overlaps in equations~\eqref{eq:quantum_Prob} and \eqref{eq:classical_Prob}.

\section*{Derivation of the quantum Fisher information}\label{sec4}

The quantum Fisher information (QFI) carried by a probe state $\ket{\psi(x)}$ about a parameter $x$, encoded in the state by the unitary evolution $\hat{U}(x)$ is given by \cite{giovannetti2011advances, demkowicz2015quantum, barbieri2022optical}

\begin{equation}
    F_Q(\ket{\psi(x)})=4\Delta^2\hat{H}\,,\qquad \hat{H}=i\frac{d\hat{U^{\dagger}}(x)}{dx}\hat{U}(x)\,,
\end{equation}
where $\hat{H}$ is the generator of the unitary $\hat{U}$ and its variance $\Delta^2\hat{H}=\langle \hat{H}^2 \rangle_\psi-\langle \hat{H} \rangle^2_\psi$ is taken with respect to the input state. In our case, where $z$ is the longitudinal displacement, the translation operator $ \hat{U}=e^{i\hat{P}_z z/\hbar}$ is the unitary of interest, yielding

\begin{equation}
    F_Q(\ket{\Psi(z)})=4\frac{\Delta^2\hat{P}_z\big\vert _\Psi}{\hbar^2}\,.
\end{equation}
Thus, to obtain the QFI we need an expression for the variance of the longitudinal momentum operator, which we provide here for the radial N00N state \eqref{eq:N00N-state}.

We start by using equations \eqref{eq:momentum-operator} and \eqref{eq:N00N-state} to obtain

\begin{equation}
\begin{aligned}
    \langle\Psi(z)\vert \hat{P}_z\vert \Psi(z)\rangle &={\hbar}\int\int {k_z}(\boldsymbol{\kappa})\, \langle \Psi(z)\vert \hat{a}^\dagger(\boldsymbol{\kappa})\hat{a}(\boldsymbol{\kappa})\vert \Psi(z)\rangle d^2\boldsymbol{\kappa}\\
&={\hbar}\sum_{ij}\int\int{k_z}(\boldsymbol{\kappa})F_i^*(\boldsymbol{\kappa})F_j(\boldsymbol{\kappa})\, \langle \Psi(z)\vert \hat{a}_i^\dagger\hat{a}_j\vert \Psi(z)\rangle d^2\boldsymbol{\kappa}\\
&=\frac{N\hbar}{2}\left( \langle k_z\rangle_p+\langle k_z\rangle_{p^\prime} \right)\,,
\end{aligned}
\label{eq:N00N-Pavg}
\end{equation}
where we used

\begin{equation}
    \left(\langle N\vert _{0p} - \langle N\vert _{0p^\prime}\right)\sum_{ij}\hat{a}^\dagger_i\hat{a}_j \left(\vert N\rangle_{0p}-\vert N\rangle_{0p^\prime}\right)=N\delta_{ip}\delta_{jp} +N\delta_{i{p^\prime}}\delta_{j{p^\prime}}\,,
    \end{equation}
and

\begin{equation}
    \langle k_z\rangle_p=\int\int{k_z}(\boldsymbol{\kappa})F_p^*(\boldsymbol{\kappa})F_p(\boldsymbol{\kappa})d^2\boldsymbol{\kappa}\,,
\end{equation}
is the classical average of the longitudinal wave vector $k_z$ for the mode $p$. Similarly, we obtain that 

\begin{equation}
    \langle\Psi(z)\vert \hat{P}_z^2\vert \Psi(z)\rangle=\frac{\hbar^2}{2}\left[ (N^2-N) \left(\langle k_z\rangle_p^2+\langle k_z\rangle_{p^\prime}^2\right) +  N\left(\langle k_z^2\rangle_p+\langle k_z^2\rangle_{p^\prime}\right)\right]\,.
\end{equation}
The resulting QFI is, therefore

\begin{equation}
F_Q(\ket{\Psi(z)})=4\frac{\Delta^2\hat{P}_z\big\vert _\Psi}{\hbar^2 }={2N} \left( \Delta^2 k_z\vert _{p}+\Delta^2 k_z\vert _{p^\prime}\right)+{N^2}\left(  \langle k_z\rangle_p-\langle k_z\rangle_{p^\prime}\right)^2\,,
\label{eq:QFI-raw}
\end{equation}
where $\Delta^2k_z\big\vert _p=\langle k_z^2\rangle -\langle k_z\rangle^2$, and equivalently for $p^\prime$.

To obtain further insight on the QFI, we swap the modes $p$ and $p^\prime$ to the Hermite-Gaussian (HG) modes HG$_{mn}$ and HG$_{m^\prime n^\prime}$, respectively.
In the HG basis, the average and variance of the longitudinal wave vector have the simple expressions 

\begin{equation}
\Delta^2 k_z\vert _{p}=\frac{1}{8z_r^2}\left[ S_2+S +2 \right]\,,\qquad \langle k_z\rangle_p =k-\frac{1}{2z_r}(S+1)\,,
\label{eq:dispersion-HG}
\end{equation}
resulting in

\begin{equation}
F_Q(\ket{\Psi(z)})=\frac{N}{4z_R^2}\left[S_2 +S_2^\prime +S +S^{\prime }+4 \right] +\frac{N^2}{4}\left(\frac{d}{dz}\Delta\Phi_G\Big\vert _{z=0}\right)^2\,,
\label{eq:QFI-HG}
\end{equation}
where $S=m+n$ is the mode order, $S_2=m^2+n^2$ (with equivalent expressions for $S^\prime$ and $S_2^\prime$), and $\Delta\Phi_G(z)=(S-S^\prime)\tan^{-1}(z/z_r)$ is the Gouy phase difference between the modes. 
The QFI is $z$-independent, consistent with the self-similarity of the angular spectrum upon propagation. It is worth noting that the second term of the QFI remains unaltered in the LG basis, while a closed form for the first term can be obtained by decomposing the LG modes in the HG basis \cite{beijersbergen1993astigmatic}. 

We note that the QFI is comprised of two fundamentally distinct terms. The first, standard-quantum limited (proportional to $N$), increases monotonically with the indices of the modes $p$ and $p^\prime$, and thus with dimension of the state space. We hypothesize that this term carries information about the full distribution of photons in the transverse plane, such that it's retrieval cannot be fully achieved by interferometric measurements alone. On the other hand, the second term displays Heisenberg-limited scaling (proportional to $N^2$) and increases with the slope of the Gouy phase difference at the focus

\begin{equation}
    \frac{d}{dz}\Delta\Phi_G\Big\vert _{z=0}=\frac{(S-S^\prime)}{z_r}\,.
\end{equation}
Equation  \eqref{eq:QFI-HG}, therefore, suggests that measurements of the quantum Gouy phase with N00N states can reach quantum-optimal sensitivity in the estimation of longitudinal displacements. 

To support our claim, we also calculate the classical Fisher information using the modeled measurement probability \eqref{eq:quantum_Prob} without the wavefront curvature. Assuming $\vert A_p\vert =\vert A_{p^\prime}\vert =A$ independent of $z$, around the region of interest $z=0$, we obtain \cite{polino2020photonic}

\begin{equation}
\begin{aligned}
    F(z)&=\sum_{i=1,2}\frac{1}{P_i}\left\vert \frac{\partial P_i}{\partial z}\right\vert ^2
    =F_Q[\mathcal{O}(N^2)]\left[\frac{4P_{max}\cos^2\left(N\delta p\tan^{-1}(z/z_R)\right)}{1-P_{max}\sin^2\left(N\delta p\tan^{-1}(z/z_R)\right)}    \right]\,,
\end{aligned}
\label{eq:CFI}
\end{equation}
where $P_{max}=2A^{2N}$, $F_Q[\mathcal{O}(N^2)]$ is the QFI term proportional to $N^2$, $P_1$ is given by \eqref{eq:quantum_Prob} and $P_2=1-P_1$. The maximum value of the Fisher information is obtained at the focus, and it's given by

\begin{equation}
    F(0)=4P_{max}F_Q[\mathcal{O}(N^2)]\,.
    \label{eq:CFI_focus}
\end{equation}
The above equation shows that the proposed measurement setup is indeed sensitive to the Heisenberg-limited part of the QFI, but does not capture any of the information contained in the SQL part. It is worth noting that equations \eqref{eq:CFI} and \eqref{eq:CFI_focus} are valid in the limit where the SMF mode is much smaller than the input modes, such that the contribution from the wavefront curvature to the measurement probability is negligible. In this case, the high losses yield $A \ll 1$ and the Fisher information is far from reaching its quantum bound. 

\section*{Experimental details}\label{sec5}

\begin{figure} [htb]
    \centering
    \includegraphics[width = \textwidth]{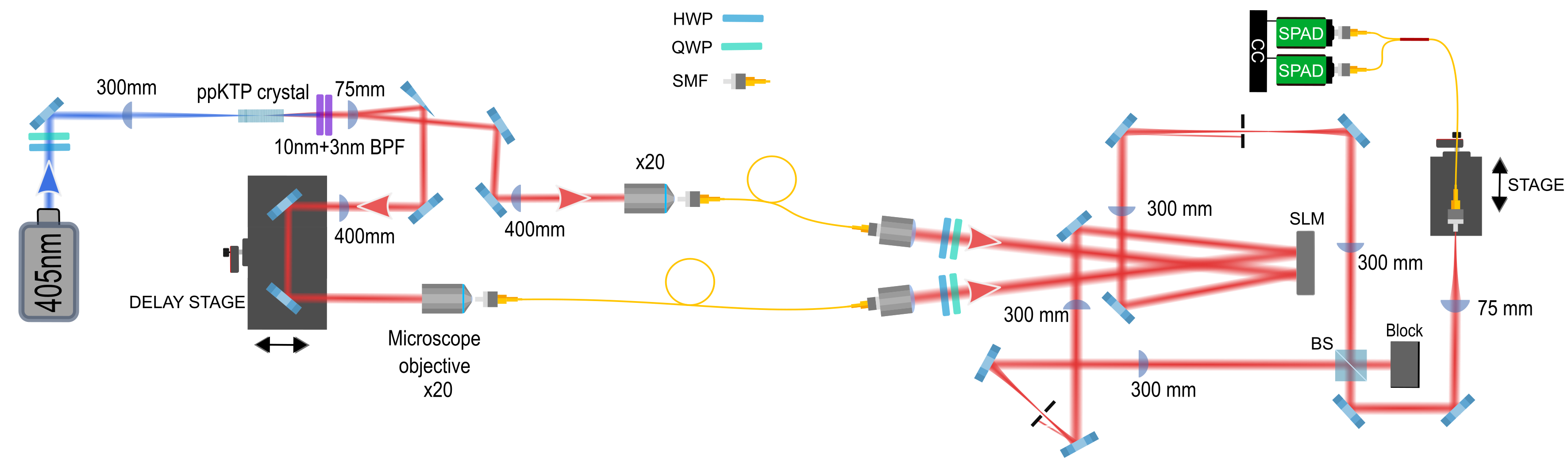}
    \caption{A detailed drawing of the experimental system. The photon pair source (left) generates photon pairs which are split and coupled into SMF's. The source also includes a delay stage which can tune the temporal distinguishability of the photon pair. The photons are sent to the spatial mode manipulation setup using the SMF's after which they are sent onto two separate regions of a spatial light modulator (SLM). After the photons have been strucutred at the SLM they are imaged with 4f systems to roughly one focal length away from the 75~mm lens which is used to focus the photons onto a SMF placed on a stage. 
    Within these identical 4f systems we place a beamsplitter (BS) which is used to probabilistically combine the photons into the same beam path.
    The half wave plates (HWP) and quarter wave plates (QWP) are used to facilitate optimum efficiency of the polarization dependent SPDC process and the SLM.}
    \label{fig:detailed_setup}
\end{figure}

The experimental setup consists of a photon pair source and the spatial mode manipulation part (see Fig.~\ref{fig:detailed_setup} for a detailed schematic).
In the photon pair source we use a 405~nm continuous-wave pump laser focused down to a 12~mm long, periodically poled nonlinear crystal made out of potassium titanyl phosphate (ppKTP).
The pump laser had a slightly astigmatic focus with an approximately 67~$\mu$m Gaussian beam waist in the crystal.
In the crystal, some of the 405~nm photons go through spontaneous parametric downconversion (SPDC) which is type 0 phase matched.
The frequencies of the two emerging 810~nm photons were made degenerate by tuning the phase matching by controlling the temperature of the crystal.
After the SPDC, the pump laser is filtered out using two bandpass filters (BPF), where the first one has a 10~nm bandpass and the second one a 3~nm bandpass around 810~nm. 
The first filter is used to remove the pump laser and the second filter is used to tune the spectral properties of the photon pair.
To split each pair of photons, we use their momentum anti-correlations by placing a lens one focal length away from the crystal after which we place a D-shaped mirror one focal length behind the lens.
The lens then performs an optical Fourier transform and the momentum anti-correlations of the photons are used to split each pair.
After splitting the photons, one of them is sent through a delay stage which controls the temporal indistinguishability of the two photons.
The photons are then coupled into separate SMF's placed on coupling stages (xyz) using a lens and a microscope objective. \\

\indent The photons then exit the SMF's after which they are collimated and sent through waveplates which align their polarizations such that the polarization sensitive SLM operates at optimum efficiency.
The SLM used was a Holoeye Pluto 2 and it was wavefront corrected using the method described in \cite{jesacher2007wavefront}.
The two photons are then modulated on two separate regions of the SLM, using mode carving which is a technique where amplitude and phase modulation can be performed on a single phase-only hologram \cite{bolduc2013exact}.
In the process, a transverse field structure $\vert u_A(x,y)\vert e^{-\textrm{i}\Phi(x,y)}$ is carved onto the first diffraction order of the hologram, out of a larger Gaussian input beam.
Quite often the effects of the initial Gaussian structure of the input beam are removed by making the Gaussian large enough to have an effectively flat amplitude distribution in the area of the hologram.
However, to produce the best possible modes at the beam radius required for our measurements, we removed the structure of the initial Gaussian by generating a hologram for the field $\frac{\vert u_A(x,y)\vert }{u_0(x,y;w_{\textrm{in}})}e^{-\textrm{i}\Phi(x,y)}$ using the same mode carving technique.
This effectively removes the structure of the incident Gaussian $u_0(x,y;w_{\textrm{in}})$ in the first diffraction order, and we only need to measure the radius $w_{\textrm{in}}$ of the beam and do not have to worry about making the incident Gaussian much larger than the hologram. 
However, we note that the incident Gaussian still needs to be slightly larger than the hologram for optimal performance.
For more information on this added Gaussian correction see the supplementary of \cite{plachta2022quantum}. \\
\indent At the SLM, we structure one of the photons in the p-mode superposition $\frac{1}{\sqrt{2}}[u_{00}(\vek{\rho},0) - e^{\textrm{i}\theta_1}u_{0p^\prime}(\vek{\rho},0)]$ and the other one in the superposition $\frac{1}{\sqrt{2}}[u_{00}(\vek{\rho},0) + e^{\textrm{i}\theta_2}u_{0p^\prime}(\vek{\rho},0)]$.
Here the phases $\theta_1$ and $\theta_2$ correspond to small corrections that need to be made due to slight imperfections in the imaging.
Due to these same imperfections, we employed additional lens terms on one of the holograms to match the focal points of the two beams as closely as possible.
After structuring, both photons go into identical 4f imaging systems and they are also sent into the same beam path, probabilistically, using a 50:50 beamsplitter.
Once in the same beam path the photons can bunch into the radial mode N00N state, as long as they are indistinguishable in every degree of freedom, which we verified in our experiments as shown below.
After this, the photons are focused down using a 75~mm focusing lens on to a SMF (Thorlabs 780HP FC/PC) placed on a translation stage that scans the SMF through the focus using a computer controlled piezo actuator (Thorlabs PIA13).
The final SMF is placed on a coupling stage (xyz) connected to a mount which can control the tip and tilt of the SMF.
After the photons are coupled into the SMF, we split them using a 50:50 fiber beamsplitter and sent them into separate single-photon avalanche diodes (SPAD; laser components COUNT T) from which we post-selected on coincident detections of two photons using a coincidence counter (CC; IDQ ID900). 
The accidental coincidences were removed from all of the data.
We calculated the accidental rates as $R_1R_2\tau$, when $R_i$ correspond to the measured single photon rates in each detector and $\tau = 1$~ns is the coincidence window used.\\
\indent In the measurements we initially set the stage as far away from the focus as possible and moved it closer to the lens, scanning the focus in discrete steps.
When calculating the distance between subsequent measurement points, we used the typical step size provided for the piezo actuator (20~nm per piezo step).
However, the manufacturer states that this step size can vary up to 20~\% depending on the component variance, change of direction, and application conditions.
Thus we tried to keep the conditions of the lab consistent while only scanning the piezo in one direction for all of the measurement.\\
\indent For the measurements with a laser, an 810~nm continuous-wave laser was used and we only required one input hologram.
We also replaced the coincident detection scheme with a power meter before and after the setup.
The results were then calculated as the power measured after the last SMF normalized by the power measured before the laser was coupled out of the input SMF.
The reference power was measured continuously by splitting the input light field using a fiber beamsplitter.
Additionally, when structuring the laser field into the superposition $\frac{1}{\sqrt{2}}[u_{00}(\vek{\rho},0) - u_{0p^\prime}(\vek{\rho},0)]$ at the measurement fiber, we had to take into account the differing Gouy phases of odd and even order modes.
Hence, when $p$ was an odd integer we had to generate the field $\frac{1}{\sqrt{2}}[u_{00}(\vek{\rho},0) + e^{\textrm{i}\theta_1}u_{0p^\prime}(\vek{\rho},0)]$ instead of $\frac{1}{\sqrt{2}}[u_{00}(\vek{\rho},0) - e^{\textrm{i}\theta_1}u_{0p^\prime}(\vek{\rho},0)]$ to compensate for the $\pi$ phase difference in Gouy phase when performing the optical Fourier transform of the field. \\

\begin{figure}[htb]
    \centering
    \includegraphics[width = 0.75\textwidth]{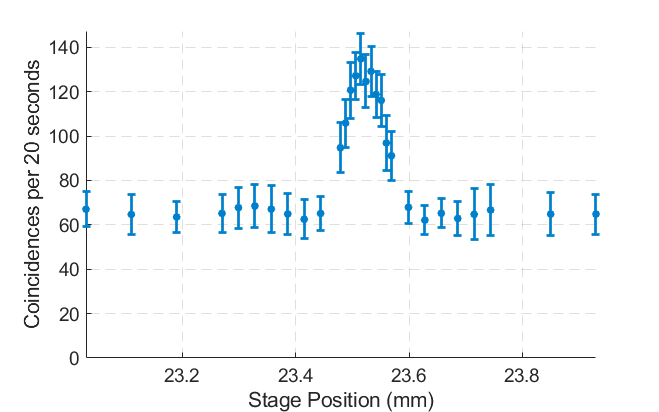}
    \caption{Measured photon bunching curve with two photons in orthogonal radial mode superpositions. Accidental coincidences have been removed from the data and the errorbars ($\pm1$ standard deviations) are calculated from 21 repetitions of the measurement. The change in stage position should be multiplied by two to arrive at the effective path length change (see Fig.~\ref{fig:detailed_setup} for the configuration of the mirrors on the delay stage). The results show that the rate at which the photon pairs couple into the SMF roughly doubles when the photons bunch into a radial mode N00N state.}
    \label{fig:Bunching}
\end{figure}
\indent To verify photon bunching in to radial mode N00N states, we set the final SMF close to the focus of the field and generated a radial mode N00N state $\frac{1}{\sqrt{2}}[\ket{2,0}_{0,2} + \ket{0,2}_{0,2}]$ by individually structuring the photons into the superpositions $\frac{1}{\sqrt{2}}[u_{00}(\vek{\rho},0) - \textrm{i}u_{02}(\vek{\rho},0)]$ and $\frac{1}{\sqrt{2}}[u_{00}(\vek{\rho},0) + \textrm{i}u_{02}(\vek{\rho},0)]$, as described earlier.
We then scan the delay stage in the source to vary the distinguishability of the photons.
From the results of the scan (Fig.~\ref{fig:Bunching}) we see that, as expected, the rate of photon pairs coupling into the SMF roughly doubles when the photons are made indistinguishable.

\bibliography{Biblio_30032022.bib}